\documentclass[jcp,aps,twocolumn,showpacs,supergroupedaddress,epsfig,amsmath,amssymb,eqsecnum]{revtex4}
\usepackage{epsfig,amsmath,amsthm,amssymb,bm,epsf,graphics,psfrag,verbatim,subfigure,framed}
\usepackage[makeroom]{cancel}
\usepackage[usenames,dvipsnames]{color}
\usepackage{empheq}
\usepackage{bbm}
\usepackage{mathrsfs}
\usepackage{amsfonts}
\usepackage{color}
\usepackage{pbox}
\def\nv{{\bf n}}

\newcommand{\be}{\begin{equation}}
\newcommand{\ee}{\end{equation}}
\newcommand{\ba}{\begin{eqnarray}}
\newcommand{\ea}{\end{eqnarray}}
\newcommand{\bw}{\begin{widetext}}
\newcommand{\ew}{\end{widetext}}
\newcommand{\Tr}{{\rm{Tr}\,}}

\newcommand{\rv}{{\bf r}}

\newcommand{\yv}{{\bf y}}

\newcommand{\cv}{{\bf c}}

\newcommand{\Rv}{{\bf R}}

\newcommand{\bing}[1]{\textcolor{black}{#1}}

\begin{document}

\title{A multiscale approach to liquid crystal nematics via statistical field theory}
\author{Bing-Sui Lu$^1$}
\email{binghermes@gmail.com,bslu@ntu.edu.sg}
\affiliation{$^{1}$ Division of Physics and Applied Physics, School of Physical and Mathematical Sciences, Nanyang Technological University, 21 Nanyang Link, 637371 Singapore.
}
\date{\today}

\begin{abstract}
We propose an approach to a multiscale problem in the theory of \bing{thermotropic uniaxial nematics} based on the method of statistical field theory. This approach enables us to relate the coefficients $A$, $B$, $C$, $L_1$ and $L_2$ of the Landau-de Gennes free energy for the isotropic-nematic phase transition to the parameters of a molecular model of \bing{uniaxial nematics}, which we take to be a lattice gas model of nematogenic molecules interacting via a short-ranged potential. We obtain general constraints on the temperature and volume fraction of nematogens for the Landau-de Gennes theory to be stable against molecular orientation fluctuations at quartic order. \bing{In particular, for the case of a fully occupied lattice, we compute the values of the isotropic-nematic transition temperature and the order parameter discontinuity predicted by (i)~a continuum approximation of the nearest-neighbor Lebwohl-Lasher model and (ii)~a Lebwohl-Lasher-type model with a nematogenic interaction of finite range. We find that the predictions of (i) are in reasonably good agreement with known results of MC simulation. }
\end{abstract}

\maketitle

\section{Introduction}

The modeling of liquid crystals is a rich and fascinating subject~\cite{degennes-prost,chandrasekhar,stephen1974,gramsbergen1986,vertogen-dejeu,ball2010,onsager1949,maier-saupe1958,maier-saupe1959,maier-saupe1960,kobayashi1971,lebwohl-lasher1972,gay-berne1981,berardi1995,berardi1998,mcmillan1971,mcmillan1972,mcmillan1973,chu-mcmillan,ronis1980,cotter1976,petrone1989a,petrone1989b,luo2012,kralj2014,kusumaatmaja2015,svensek2010,svensek2013,dean2012,han2015,pasini2000a,pasini2000b,gelbart-baron1977,cotter1977}. The models span a hierarchy of levels of coarse-graining approximation and attention to the details of interactions. At the upper end of the hierarchy is the phenomenological approach of the Landau-de Gennes type~\cite{degennes-prost,gramsbergen1986,ball2010,luo2012,kralj2014,kusumaatmaja2015}, which involves the construction of a free energy using invariant combinations of an order parameter and its gradients, the order parameter being a coarse-grained quantity that reflects the amount of liquid crystalline order there is at the mesoscopic level. Such a free energy also involves several adjustable coefficients that have to be fitted to experiment, whose values are not known {\it a priori}. The Landau-de Gennes free energy can be expressed as~\cite{gramsbergen1986} 
\ba
F_{{\rm LdG}} &\!=\!& \! \int \! \frac{d^3\rv}{a^3} \, 
\Big[ \frac{1}{2} L_1 (\partial_a Q_{bc})^2 + \frac{1}{2} L_2 (\partial_a Q_{ab})^2
\nonumber\\
&&+ 
\frac{A}{2} \Tr Q^2 
- \frac{B}{3} \Tr Q^3 + \frac{C}{4} (\Tr Q^2)^2
\Big], 
\ea
where $Q$ is a second-rank, traceless and symmetric tensor, directly related to the degree of alignment of liquid crystal nematics, and $a$ is a microscopic lengthscale related to the dimensions of the molecule. 

At the other end of the hierarchy are the molecular models of liquid crystals (LC), such as the Onsager model~\cite{onsager1949,han2015} for lyotropic LC or the Maier-Saupe~\cite{maier-saupe1958,maier-saupe1959,maier-saupe1960}, Lebwohl-Lasher~\cite{lebwohl-lasher1972,pasini2000a,pasini2000b} and Gay-Berne models~\cite{gay-berne1981,berardi1995,berardi1998} for thermotropic LC. These models have for their parameters certain details of the interacting molecules, which can include the aspect ratio, the orientations and positions of the molecules, and the form of the interaction potential between each pair of molecules. In choosing the interaction potential, there is consensus that the relevant factors giving rise to nematic liquid crystalline ordering involve suitably anisotropic generalizations of the excluded-volume interaction and/or the van der Waals interaction~\cite{gelbart-baron1977,cotter1977}. On the other hand, models also differ in the scale of the relative importance of anisotropy they assign to excluded-volume and van der Waals interactions. For example, the theory of Onsager~\cite{onsager1949} assumes that nematic ordering is  driven by the anisotropy of hard rods' excluded volumes, whereas the model of Maier and Saupe~\cite{maier-saupe1958,maier-saupe1959,maier-saupe1960} assumes that the driving agent is the anisotropy in the van der Waals interaction. Between these two extremes, there is also the so-called generalized van der Waals theory, which takes the anisotropies present in both the excluded volumes of the molecules and the van der Waals interaction into consideration~\cite{gelbart-baron1977,cotter1977}. 

Beyond the individual modeling of liquid crystals, one can also study the relations between different theories of LC, especially theories on different levels of the coarse-graining hierarchy~\cite{chandrasekhar,han2015,sauerwein-oliveira2016}, and our paper purports to address such a problem. This problem appears not to have received as much attention as the modeling of LC. An example of such a problem would be to establish quantitative relations between the coefficients of the Landau-de Gennes (LdG) theory and the parameters of a molecular model. Establishing such relations is of relevance to the LC community, one of the motivations being that it would make the LdG theory more predictive. There are different ways by which such relations can be established, the differences being primarily differences of the molecular model and approximation scheme adopted. One approximation scheme is to begin with a certain molecular model of LC and apply the mean-field approximation at the level of the interaction Hamiltonian~\cite{chandrasekhar}. One then computes the corresponding partition function and free energy, matching the coefficients of the free energy to those of the LdG free energy. Such an approach has been applied, for example, to a molecular model of biaxial bricks~\cite{sauerwein-oliveira2016}. A variant of the mean-field approach is to express the the entropy of the molecules and their average pairwise interaction energy in terms of the distribution function for the molecular positions and orientations, determining this function self-consistently via some form of closure at the mean-field level. Such an approach has been applied, for example, to relate the Onsager theory~\cite{onsager1949} to the LdG theory~\cite{han2015}. In mean-field approaches the mean-field approximation is typically applied to the Hamiltonian, i.e., replacing a Hamiltonian of the form $q^2$, where $q$ is a fluctuating field variable, with a Hamiltonian of the form $q \, f(\langle q \rangle)$, where $\langle q \rangle$ is the mean field and $f$ is some function of the mean field, so that the ordeal of summing over $q$ in the partition function is much simplified as one is now dealing with a term that is linear rather than quadratic in $q$. This also implies that the corresponding LdG coefficients obtained by the molecular calculation are given by their {\em mean-field} values and neglect corrections from the correlations of {\em fluctuations}. In addition to mean-field approaches, there is also the approach of density functional theory, which views the free energy as being recoverable if all its direct correlation functions (which are the moments of the free energy with respect to particle density) are known~\cite{hansen-mcdonald}.  Such an approach was used to relate the elastic constants of the Frank-Oseen theory to the single-nematogen orientation distribution function~\cite{singh1986a,singh1986b,singh1987} and has also been used to establish formal relations between the LdG theory of the isotropic-nematic transition and the single-nematogen orientation distribution function~\cite{singh1984}. 

In view of the limited number of methods of relating the mesoscopic (or continuum, i.e., LdG) and the microscopic (or molecular) levels of description in LC theory, and the limitations of the mean-field approximation, it may be of interest to explore a third approach, namely, that of field theory~\cite{doi-edwards,fredrickson,edwards-lenard1962,lu2012,lu2013,polyakov,altland-simons}, which leads to the inclusion of the effects of molecular fluctuation correlations in the computed values of the LdG coefficients. The present paper explores this approach, applying it to a molecular lattice model of Lebwohl-Lasher type~\cite{lebwohl-lasher1972}. Such an approach would also be a step towards the realization of a programme of calculation envisioned in Ref.~\cite{degennes-prost}, whereby ``one should start from a microscopic Hamiltonian" and ``calculate whatever thermodynamic quantity is needed from the partition function $Z$", and that it should be ``possible to construct a free energy $F(\langle Q \rangle)$, the minima of which do rigorously define equilibrium states", where $F(\langle Q \rangle) = -k T \log Z(\langle Q \rangle)$, and the integrations in $Z$ ``are performed with a constant average order parameter $\langle Q \rangle$." 

Another motivation for using the field-theoretic approach (in particular, the Hubbard-Stratonovich transformation~\cite{stratonovich1958,hubbard1958,siegert1960}, which we shall describe shortly) to study the microscopic-mesoscopic relation comes from the fact that it appears to be better known in other domains of physics, but seems much less studied in the domain of liquid crystals. For example, in the theory of magnetism, field theory is used to relate the mesoscopic coefficients of the $\phi^4$ theory to the molecular parameters of the Ising model~\cite{polyakov,altland-simons}. In the theory of electrolytes, field theory is used to relate a microscopic Coulomb model of interacting ions to Poisson-Boltzmann (PB) theory, which is a continuum description~\cite{podgornik-zeks1988,podgornik1989a,podgornik1989b,netz-orland2000b,netz2001}. In the domain of liquid crystals, we are only aware of \bing{a couple of works~\cite{liu-fredrickson1993,drossinos1986} which adopt a field-theoretic approach. 
In Ref.~\cite{liu-fredrickson1993} the authors apply the field-theoretic approach to study the phase behavior of semiflexible polymer solutions and blends, starting from a model that includes both an anisotropic interaction that favors the alignment of polymer segments and an isotropic interaction that can drive the demixing of polymer and solvent. 
In Ref.~\cite{drossinos1986}} the authors analyze the phenomenological theory for smectics in terms of the molecular parameters of the Ronis-Rosenblatt model~\cite{ronis1980}, neglecting however to obtain the full coefficients of the nematic theory. 
Our paper endeavors to relate the phenomenological coefficients of the \bing{theory of thermotropic uniaxial nematics} to a Lebwohl-Lasher-type lattice model~\cite{footnote-lattice}, and can be regarded as being complementary to the aforementioned work. 

In what follows, we qualitatively describe the field-theoretic method which we use, relegating the mathematical details to subsequent sections. This method begins by specifying a microscopic Hamiltonian that can be expressed in terms of a quadrature of some collective field. One then performs a Hubbard-Stratonovich (HS) transformation~\cite{stratonovich1958,hubbard1958,siegert1960} on the partition function, which is an exact transformation that introduces an {\em auxiliary} field conjugate to the original collective field, and causes the collective field to appear in a linearly coupled form, rendering tractable the task of ensemble averaging over the original fluctuating degrees of freedom (which are the molecular orientation and lattice site occupation number). We then show that performing an ensemble average of the collective field over the original degrees of freedom is equivalent to computing the expectation value of the auxiliary field, which corresponds to the $\langle Q \rangle$ of the LdG theory. The resulting free energy with the log trace term can be systematically expanded as a power series of $Q$, the coefficients of the terms being combinations of the parameters of the microscopic model and the concentration of nematogens. Identifying this series (which we truncate at quartic order) with the LdG free energy then enables the determination of the coefficients $A$, $B$ and $C$ as well as the elastic constants $L_1$ and $L_2$ in terms of the molecular parameters. Via such a procedure, we will also discover that a certain condition has to hold in order to ensure the positivity of the quartic term (and hence the stability of the LdG theory), namely, that the volume fraction of nematogens has to be sufficiently large. 

The plan of our paper is as follows. In Sec.~II, we explicate the model that we adopt, which is a modified version of the Lebwohl-Lasher model. In Sec.~III, we construct the partition function and effect the Hubbard-Stratonovich transformation, which enables us to transition from a picture of fluctuating molecular orientations and lattice site occupation to a picture of fluctuating order parameter field. In Sec.~IV, we derive the effective Hamiltonian and show that its quartic expansion can be identified with the Landau-de Gennes free energy. This paves the way to our results in Sec.~V, where we identify the phenomenological coefficients of the Landau-de Gennes theory in terms of the molecular parameters of our modified Lebwohl-Lasher model. \bing{In Sec.~VI, we look at the case of a fully occupied lattice, and obtain values for the isotropic-nematic transition temperature and the order parameter discontinuity that are predicted by the two following models: (i)~a continuum approximation of the nearest-neighbor Lebwohl-Lasher model and (ii)~a model of Lebwohl-Lasher type that involves nematogenic interactions of finite range.}

\section{Modified Lebwohl-Lasher model}
\begin{figure}[h]
\centering
  \includegraphics[width=0.38\textwidth]{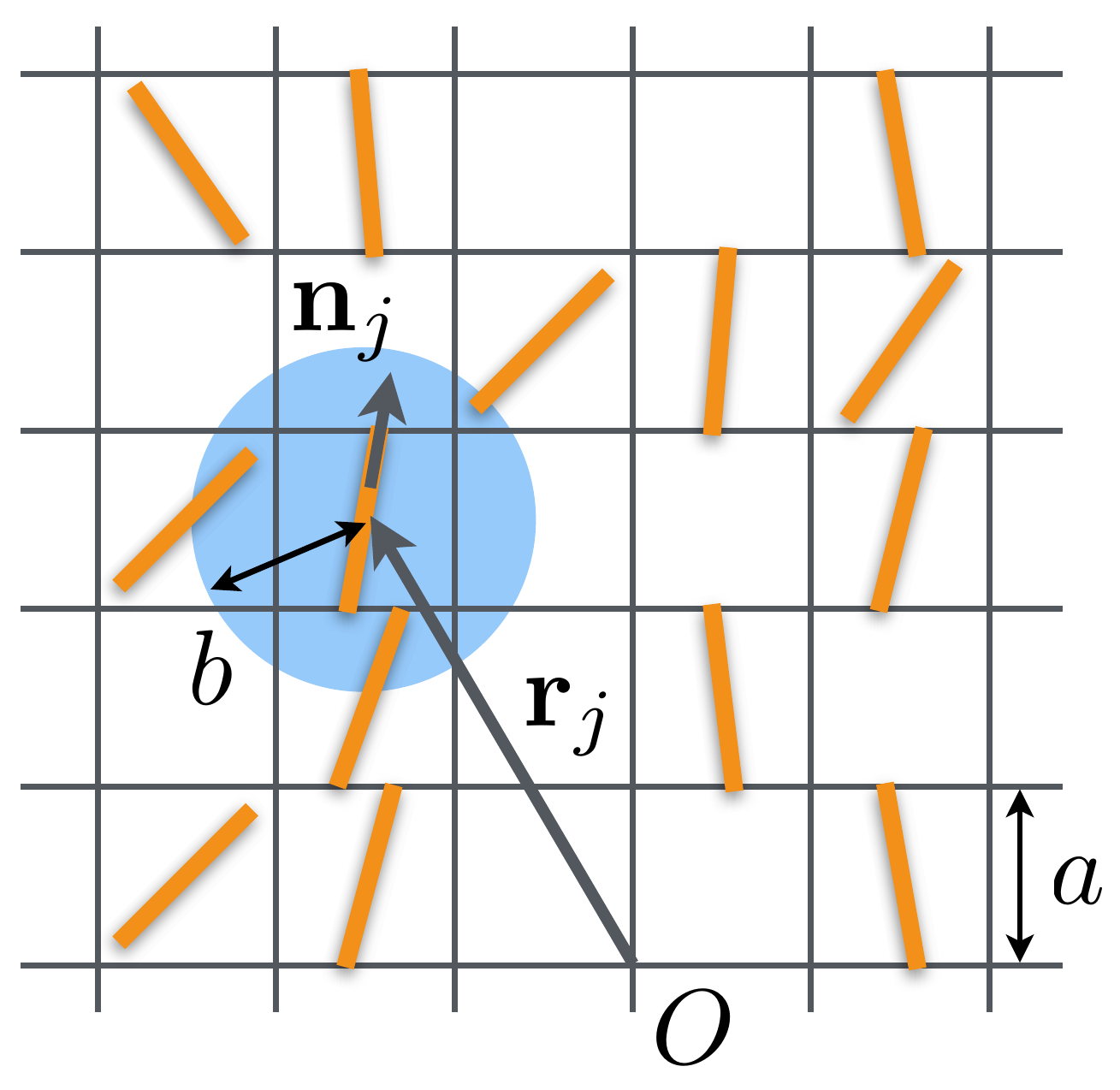}
  \caption{Modified Lebwohl-Lasher model: a two-dimensional slice of the three-dimensional cubic lattice on which nematogens (represented by orange-colored rods) reside. The $j$th rod has a direction parallel to $\nv_j$ and a position vector $\rv_j$, inhabiting a cell that has a linear dimension $a$. A cell can either be occupied by a nematogen, or unoccupied. The circle (colored blue, radius $b$) indicates the effective range of the nematogen-nematogen interaction potential $v(\rv)$.} 
  \label{fig:lattice}
\end{figure}
We consider a lattice gas model of liquid crystal nematics, in which the nematogens reside on a three-dimensional cubic lattice. Each lattice cell has a volume $a^3$, where $a$ is the length of each side of the cell. We assume that each nematogen has a volume $a^3$, so each cell can at most be occupied by one nematogen. We denote the occupation number of the $j$-th cell by $s_j$, where $s_j=0,1$, its position vector by $\rv_j$, and the orientation of the nematogen in the cell is represented by a unit vector $\nv_j$. The nematogen-nematogen interaction is described by a Lebwohl-Lasher-type model~\cite{lebwohl-lasher1972}, which has a Hamiltonian given in dimensionless form by
\be
\beta H_{LL} = -\frac{1}{2} \sum_{i,j=1}^{N} v(\rv_i-\rv_j) \Big( \left( {\bf n}_i \!\cdot\! {\bf n}_j \right)^2 - \frac{1}{3} \Big)
\ee
where $v(\rv_i - \rv_j) \equiv \beta J_{ij}$, and $J_{ij}$ is the interaction between nematogens at positions $\rv_i$ and $\rv_j$. 
Here the Latin indices $i, j = 1, \dots, N$ label nematogens. \bing{Our model is thus a modification of the original Lebwohl-Lasher (LL) model in two respects, viz., in the allowance we make for partial occupation of the lattice, and in that the nematogen interaction is not restricted solely to nearest-neighbor interactions. }

\section{Partition function}

The partition function is given by 
\be
\label{eq:Zorig}
Z = \prod_{j=1}^{N} \sum_{s_j=0,1} \! {\rm e}^{\beta \mu \sum_j s_j} 
\Big\langle 
{\rm e}^{\frac{1}{2}\sum_{i,j} 
v(\rv_i - \rv_j) 
\big( \left( {\bf n}_i \cdot {\bf n}_j \right)^2 - \frac{1}{3} \big) 
s_i s_j }
\Big\rangle_{n_j}
\ee
where $\mu$ is the chemical potential of each nematogen. The chemical potential can be regarded as a Lagrange multiplier that enforces the conservation of the average number of nematogens, so the partition function represents a trace over all possible configurations of $\{ s_j \}_{j=1}^{N}$ subject to the constraint that the average number of nematogens is conserved. 
The symbol $\langle \ldots \rangle_{n_j}$ denotes the average over all possible directions of the unit vector $\nv_j$, viz., 
\be
\langle \ldots \rangle_{n_j} \!\equiv\! 
\frac{d\,\Omega_j}{4\pi} \,
\delta(|\nv_j|^2-1),  
\ee 
where $d\,\Omega_j$ is an infinitesimal element of solid angle associated with the $j$th nematogen. 
By defining a collective field $q_{ab}({\bf r})$, viz., 
\be
q_{ab}({\bf r}) 
\equiv
\sum_{j=1}^{N} \left( n_{j a} n_{j b} - \frac{1}{3}\delta_{ab} \right) s_j a^3 
\delta({\bf r} - {\bf r}_j), 
\ee
where $q_{ab}(\rv)$ has the meaning of the local liquid crystalline order, we can rewrite the partition function as 
\be
\label{eq:Z}
Z = \prod_{j=1}^{N} \sum_{s_j=0,1} \! {\rm e}^{\beta \mu \sum_j s_j} \! 
\left\langle 
{\rm e}^{\frac{1}{2} \!\int\! \frac{d^3\rv}{a^3} \! \int \! \frac{d^3\rv'}{a^3} \, 
q_{ab}(\rv) v(\rv - \rv') q_{ab}(\rv')}
\right\rangle_{n_j}
\ee
We now make use of the Hubbard-Stratonovich transformation, which is based on the matrix integral identity
\be
\exp (\tfrac{1}{2} x_i K_{ij} x_j) \propto\int \! d\yv \exp(-\tfrac{1}{2} y_i K_{ij} y_j + K_{ij} x_i y_j). 
\ee
We can therefore express the partition function as 
\ba
\label{eq:Zdual}
Z &\!\!=\!\!& \int\!\mathcal{D}Q \, 
{\rm e}^{-\frac{1}{2} \!\int\! \frac{d^3\rv}{a^3} \! \int \! \frac{d^3\rv'}{a^3} \, 
Q_{ab}(\rv) v(\rv - \rv') Q_{ab}(\rv')}
\\
&&\times
\prod_{j=1}^{N} 
\sum_{s_j=0,1} \! 
{\rm e}^{\beta \mu \sum_{j} s_j}
\Big\langle
{\rm e}^{\!\int\! \frac{d^3\rv}{a^3} \! \int \! \frac{d^3\rv'}{a^3} \, 
q_{ab}(\rv) v(\rv-\rv') Q_{ab}(\rv')}
\Big\rangle_{n_j}
\nonumber
\ea
Here the integral measure $\int \!\mathcal{D}Q$ is a shorthand for the product $\prod_{\{\rv\}}\! dQ(\rv) \equiv \lim_{N\rightarrow\infty}\prod_{J=1}^{N} \! dQ(\rv_J)$, where for the purpose of calculation we regard the volume as being a cubic lattice with $N$ cells. 
We rewrite the product on the right-hand side (RHS) as
\ba
&&\prod_{j=1}^{N} 
\sum_{s_j=0,1} \! 
{\rm e}^{\beta \mu \sum_{j} s_j}
\Big\langle
{\rm e}^{\int \! \frac{d^3\rv}{a^3} \! 
\int \! \frac{d^3\rv'}{a^3} 
\, q_{ab}(\rv) v(\rv-\rv') Q_{ab}(\rv')}
\Big\rangle_{n_j}
\nonumber
\\
&=& 
\prod_{j=1}^{N} 
\left[ 
1 + \lambda \Big\langle 
{\rm e}^{\int \! \frac{d^3\rv}{a^3} v(\rv-\rv_j) Q_{ab}(\rv)(n_{ja}n_{jb}-\frac{1}{3}\delta_{ab})}
\Big\rangle_{n_j}
\right]
\nonumber\\
&=& 
{\rm e}^{\sum_j \ln 
\big[ 
1 + \lambda \big\langle 
{\rm e}^{\int \! \frac{d^3\rv}{a^3} v(\rv-\rv_j) Q_{ab}(\rv)(n_{ja}n_{jb}-\frac{1}{3}\delta_{ab})}
\big\rangle_{n_j}
\big]},
\nonumber
\ea
where $\lambda \equiv {\rm e}^{\beta\mu}$ is the fugacity of a nematogen. Next, for a dense lattice we can adopt a coarse-graining approximation in which we replace the sum over lattice cells by an integral over the entire volume enclosed by the lattice. We then obtain 
\ba
Z &\!=\!& \int\!\mathcal{D}Q \, 
{\rm e}^{-\frac{1}{2} \!\int\! \frac{d^3\rv}{a^3} \! \int \! \frac{d^3\rv'}{a^3} \, 
Q_{ab}(\rv) v(\rv - \rv') Q_{ab}(\rv')}
\\
&&\times \, 
{\rm e}^{\, \int \! \frac{d^3\rv}{a^3} \ln
\big[ 
1 + \lambda \big\langle 
{\rm e}^{\int \! \frac{d^3\rv'}{a^3} v(\rv-\rv') 
Q_{ab}(\rv')(n_{a}n_{b}-\frac{1}{3}\delta_{ab})}
\big\rangle_{n}
\big]}
\nonumber
\ea

\section{Effective Hamiltonian}

We can formally define an effective Hamiltonian $H_{{\rm eff}}$ via
\be
\label{eq:ZQ}
Z = \int \! \mathcal{D}Q \, e^{-\beta H_{{\rm eff}}},
\ee
and we shall posit that $H_{{\rm eff}}$ is equivalent to the Landau-de Gennes (LdG) free energy. Indeed, we will see shortly that the auxiliary field $Q$ has the meaning of the order parameter in the LdG theory. Through this identification, the coefficients of the LdG theory can be determined in terms of the parameters of the microscopic model -- an objective which we set out to achieve. 
Up to some constant, we can write the effective Hamiltonian as
\ba
\label{eq:HeffQ}
&&\beta H_{{\rm eff}} 
\\
&\!=\!& 
\frac{1}{2} 
\!\int \! \frac{d^3\rv}{a^3} \! \int \! \frac{d^3\rv'}{a^3} \, 
Q_{\alpha\beta}(\rv) v(\rv-\rv') Q_{\alpha\beta}(\rv') 
\nonumber\\
&&- \int \! \frac{d^3\rv}{a^3} \ln \! 
\big[ 
1 + \lambda \big\langle 
{\rm e}^{\int \! \frac{d^3\rv'}{a^3} v(\rv-\rv') \tau_{ab} Q_{ab}(\rv')}
\big\rangle_{n}
\big],
\nonumber
\ea
where $\tau_{ab} \equiv n_{a}n_{b}-\frac{1}{3}\delta_{ab}$. 
The fugacity $\lambda$ is related to the total number of nematogens, $N_{nem}$, on the lattice, which we can see as follows. Making use of Eqs.~(\ref{eq:ZQ}) and (\ref{eq:HeffQ}) and the thermodynamic relation 
\be
N_{nem} = \frac{1}{Z} \frac{\partial Z}{\partial (\beta \mu)} 
\ee
which (as mentioned earlier) reflects the fact that $\mu$ is a Lagrange multiplier that ensures that the average number of nematogens is conserved, we have 
\be
N_{nem} a^3 = \int\! d^3\rv \left\langle \frac{\lambda \langle {\rm e}^{\int \! v \,\tau_{ab} Q_{ab}} \rangle_n}{1 + \lambda \langle {\rm e}^{\int \! v \, \tau_{ab} Q_{ab}} \rangle_n} \right\rangle_Q,
\ee
where $\int (\ldots)$ is our shorthand for $\int\! d^3\rv/a^3 (\ldots)$, and $\langle \ldots \rangle_Q \equiv \int \! \mathcal{D}Q (\ldots) {\rm e}^{-\frac{1}{2} \! \int \! Q v^{-1} Q}/ \! \int \! \mathcal{D}Q \, {\rm e}^{-\frac{1}{2} \! \int \! Q v^{-1} Q}$ is the Gaussian weighted average over $Q$. The left-hand side (LHS) of the above equation depends on how the system is prepared, and once prepared, is assumed to remain constant for the entire duration, while the temperature (and correspondingly the expectation value of $Q$) can change. The fugacity $\lambda$ of each nematogen is also assumed to be constant. Thus, to determine $\lambda$, let us consider the regime of high temperature, for which $T \rightarrow \infty$ (or equivalently, $v \rightarrow 0$). The above equation then becomes
\be
N_{nem} a^3 \rightarrow \frac{V\lambda}{1+\lambda},
\ee 
where $V$ is the volume of the system. 
Further denoting the fraction of lattice cells occupied by nematogens by $\phi \equiv N_{nem} a^3/V = N_{nem}/N$, we find that 
\be
\lambda = \frac{\phi}{1-\phi}.
\ee
As we shall see soon (i.e., end of this section and App.~\ref{app:Q}), $Q$ has the meaning of the Landau-de Gennes order parameter. Anticipating this result, and observing that the LdG free energy is formulated near the isotropic-nematic phase transition, we ca
thus carry out an expansion of the effective Hamiltonian in powers of $Q$. We carry out this expansion in App.~\ref{app:Hexpansion}. To quartic order in $Q$, we have the result that the effective Hamiltonian is given by
\ba
\label{eq:nonlocal0}
&&\beta H_{{\rm eff}} 
\\
&\!=\!& 
I_1 - \frac{\lambda}{15(1+\lambda)} I_2 
- \int \! \frac{d^3\rv}{a^3} 
\Big\{ 
\frac{4 \kappa^3 \lambda}{315(1+\lambda)} \Tr Q^3 
\nonumber\\
&&\quad + 
\left[ \frac{\lambda}{\bing{567}(1+\lambda)} - \frac{\lambda^2}{450(1+\lambda)^2} \right] 
\kappa^4
(\Tr Q^2)^2 
\Big\},
\nonumber
\ea
where we have defined 
\begin{subequations}
\ba
\label{eq:I1}
I_1 &\!\equiv\!&
\frac{1}{2} 
\!\int \! \frac{d^3\rv}{a^3} \! \int \! \frac{d^3\rv'}{a^3} \, 
Q_{\alpha\beta}(\rv) v(\rv-\rv') Q_{\alpha\beta}(\rv') 
\\
\label{eq:I2}
I_2 &\!\equiv\!&
\! \int \! \frac{d^3\rv}{a^3} 
\! \int \! \frac{d^3\rv'}{a^3}
\! \int \! \frac{d^3\rv''}{a^3}
v(\rv'-\rv) v(\rv''-\rv)
\nonumber\\
&&\qquad\quad \times \, 
Q_{ab}(\rv') Q_{ab}(\rv'') 
\\
\kappa 
&\!\equiv\!& 
\int \! \frac{d^3\rv}{a^3} v(\rv).
\ea
\end{subequations}
In obtaining Eq.~(\ref{eq:nonlocal0}), we have made use of the fact that for symmetric, traceless $3 \times 3$ matrices, $\Tr Q^4 = (\Tr Q^2)^2 / 2$~\cite{footnote1}. 
In carrying out the orientation averages, we have made use of certain tensor results that we derived in Appendix~\ref{app:tensors}. 

What is the meaning of $Q$? In App.~\ref{app:Q}, we show that $Q$ has the meaning of the order parameter field in the LdG theory. We demonstrate this by showing that the expectation value of $Q$, obtained by averaging over $Q$ with a Boltzmann weight $e^{-\beta H_{{\rm eff}}}$, coincides with the ensemble average of $q$, i.e., 
\be
\langle q_{ab}(\rv) \rangle = \langle Q_{ab}(\rv) \rangle. 
\ee
Here the angle brackets have somewhat distinct meanings on the left and right sides of the equation; on the left side they denote averaging with respect to the ensemble variables $s_j$ and $\nv_j$ using the partition function in Eq.~(\ref{eq:Zorig}), whereas on the right the averaging is performed over $Q$ using the partition function in Eq.~(\ref{eq:ZQ}).

\section{Matching coefficients}

The first two terms (i.e., the ones involving $I_1$ and $I_2$) in Eq.~(\ref{eq:nonlocal0}) are nonlocal. To make contact with LdG theory, we perform a gradient expansion on those terms, whereupon we obtain (see App.~\ref{app:calc} for details) 
\ba
\label{eq:Heff-locale}
\beta H_{{\rm eff}} 
&\!=\!& 
\kappa \left( \frac{1}{2} - \frac{\lambda \, \kappa}{15(1+\lambda)} \right) 
\!\int \! \frac{d^3\rv}{a^3} \, 
\Tr Q^2
\\
&&
+ \, \eta \left( \frac{\lambda \, \kappa}{15(1+\lambda)} 
- \frac{1}{4} \right)
\! \int \! \frac{d^3\rv}{a^3} 
\partial_c Q_{ab}
\partial_c Q_{ab}
\nonumber\\
&&- \int \! \frac{d^3\rv}{a^3} 
\bigg\{ 
\frac{4 \kappa^3 \lambda}{315(1+\lambda)} \Tr Q^3 
\nonumber\\
&&\quad + 
\left( \frac{\lambda}{\bing{567}(1+\lambda)} - \frac{\lambda^2}{450(1+\lambda)^2} \right) 
\kappa^4
(\Tr Q^2)^2 
\bigg\}.
\nonumber
\ea
The coefficient of the cubic term in the Landau expansion is negative, which is in accord with expectations for elongated nematogens. 
In the above expression, we have also defined
\be
\eta \, \delta_{ab} 
\equiv\! 
\int \! \frac{d^3\Rv}{a^3}
R_{c} R_{d} \, v(\Rv), 
\ee
where a Kronecker delta appears because $v$ is isotropic. If we assume that $v$ has the form of Eq.~(\ref{eq:v-gaussian}) then 
\be
\eta = (2\pi)^{3/2} \beta J b^5/a^3.
\ee
Our expansion for $H_{{\rm eff}}$ in Eq.~(\ref{eq:Heff-locale}) coincides with the Landau-de Gennes free energy, $F_{{\rm LdG}}$, if we make the following identification: 
\begin{subequations}
\label{eq:matched}
\ba
A &\!=\!& \left( 1 - \frac{2 \lambda \, \kappa}{15(1+\lambda)} \right) \kappa \, k_{{\rm B}}T,
\\
B &\!=\!& \frac{4 \kappa^3 \lambda \, k_{{\rm B}}T}{105(1+\lambda)},
\\
C &\!=\!& \left( \frac{2\lambda^2}{225(1+\lambda)^2} - \frac{4\lambda}{567(1+\lambda)} \right) 
\kappa^4 k_{{\rm B}}T,
\\
L_1 &\!=\!& 2\eta \left( \frac{\lambda \, \kappa}{15(1+\lambda)} 
- \frac{1}{4} \right) k_{{\rm B}}T,
\\
L_2 &\!=\!& 0
\ea
\end{subequations}
From these equations, we can make the following deductions. We first note that $A$ becomes negative for $\kappa > 15(1+\lambda)/2\lambda$, which allows us to write $A = A_0 (T - T^*)$, \bing{where $A_0 \equiv \kappa k_{{\rm B}}$.} 
Here $T^*$ corresponds to the critical temperature, which has the physical meaning of the limit of supercooling of the isotropic state. Using the matched coefficients above, we can express $T^*$ in terms of the molecular parameters: 
\be
\label{eq:T*}
T^* = 2\lambda \kappa T/(15(1+\lambda)).
\ee
\bing{Here $T^*$ is actually independent of $T$ as $\kappa$ contains a factor $1/T$. From Landau-de Gennes theory, we have that $T_{ni} = T^* + B^2/(27 C A_0)$, and the discontinuity in the nematic order is $S_c = B/3C$. For the case of a fully occupied lattice ($\lambda \rightarrow \infty$) and using the matched coefficients in Eqs.~(\ref{eq:matched}), we find that 
\be
T^* = \frac{2\kappa T}{15}, \quad T_{ni} = \frac{74}{455} \kappa T, \quad S_c = \frac{5670}{1079\kappa}. 
\ee}
The coefficient $L_2$ is zero, implying that our expansion is equivalent to the one-constant approximation~\cite{degennes-prost}, which is a consequence of adopting an isotropic form for the nematogen interaction potential $v$.
\begin{figure}[h]
\centering
  \includegraphics[width=0.47\textwidth]{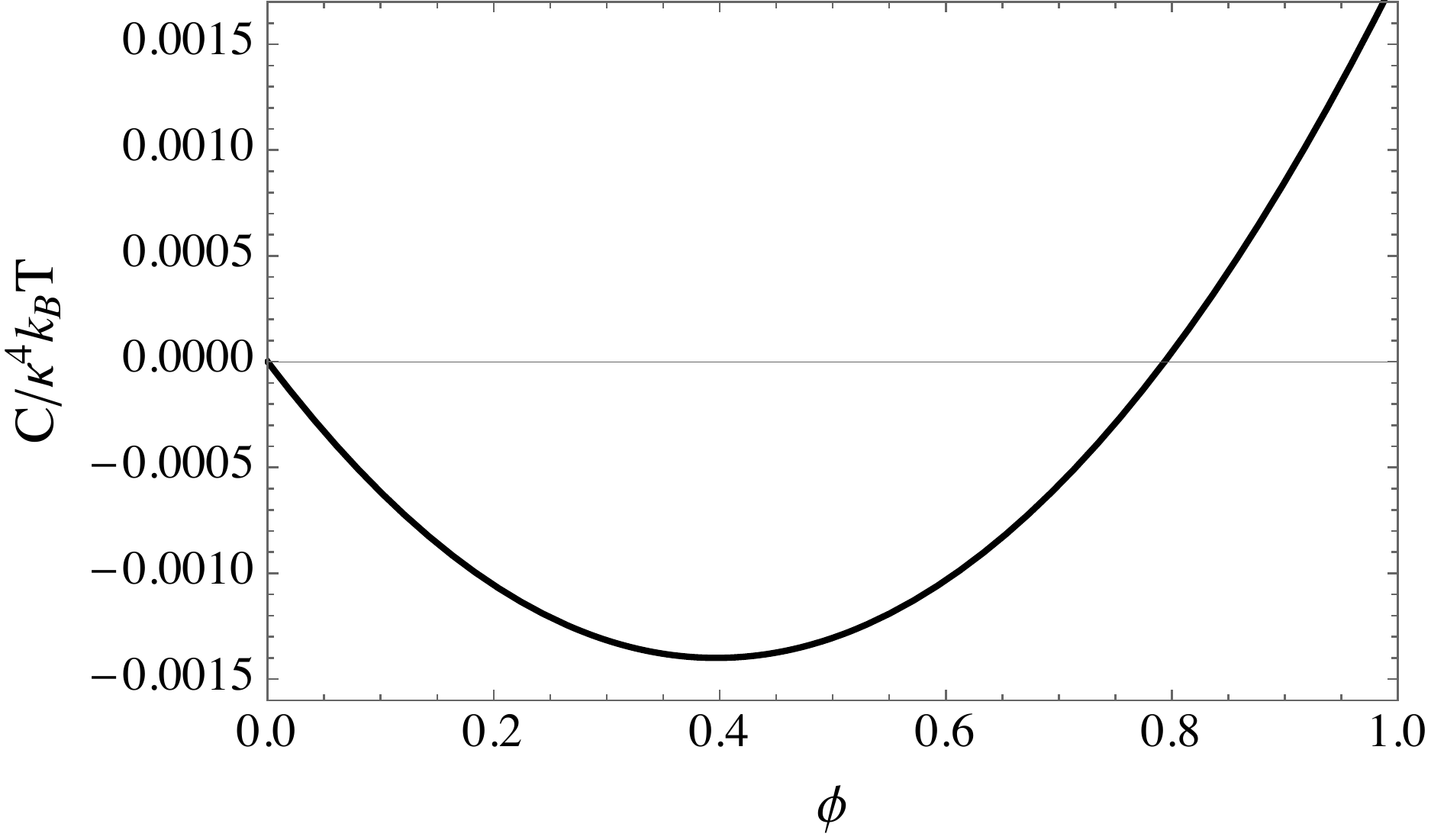}
  \caption{Behavior of the coefficient $C$ of the Landau-de Gennes theory as a function of $\phi$, the fraction of lattice cells occupied by nematogens. $C$ is positive for \bing{$\phi > 50/63$.}} 
  \label{fig:C}
\end{figure}
We also see that besides the possibility of $A$ becoming negative, $L_1$  also becomes negative
 for $\kappa > 15(1+\lambda)/4\lambda$. The sign changes of the quadratic term and quadratic gradient are well-known from field-theoretic expansions (see, e.g., Ref.~\cite{altland-simons}), and describe the limits at which the expansion to quadratic order is stable against the effects of fluctuation. 
 
In Eqs.~(\ref{eq:matched}), we observe that the parameter $C$, corresponding to the coefficient of the quartic term, is positive for \bing{$\lambda > 50/13$ (i.e., $\phi > 50/63$)}, but then changes sign and becomes negative for \bing{$0 \leq \lambda < 50/13$ (i.e., $0 \leq \phi < 50/63$).} Similar to the sign changes of the quadratic coefficients, the sign change of the quartic coefficient is also generic to field-theoretic expansions~\cite{altland-simons}, arising simply from the competition of the second and fourth moments of the fluctuation correlation function of molecular orientation in the cumulant expansion (cf. Eq.~(\ref{eq:nonlocal})). Such a possibility of a change in the sign of the quartic coefficient is absent in the more conventional, mean-field approximation applied at the Hamiltonian level. 
For a given volume fraction of nematogens, the sign provides an indication of whether the truncation of the LdG theory to quartic order is stable against  fluctuations. 
As we see from Fig.~\ref{fig:C}, such a truncation is stable only if the volume fraction of nematogens is sufficiently large. 

\bing{We may also make a qualitative comparison of our results with those found for semiflexible polymer solutions in Ref.~\cite{liu-fredrickson1993}, where polymer rigidity (besides the alignment interaction strength of polymer segments) can also drive nematic ordering. If one assumes the absence of density fluctuations, the coefficient of the quartic order term turns out to be positive for any value of the volume fraction, which is distinct from the behavior predicted by the modified LL model, in which $C$ can become negative. This suggests that the combination of both polymer rigidity and alignment interaction stabilizes the LdG theory at quartic order. The authors also found that the coefficient $L_2$ is non-zero, which can be understood as a consequence of the preference of segments to align parallel to the semiflexible backbone, whereas for models I and II $L_2$ vanishes, owing to the assumed isotropy of $v(\rv)$.}  

\section{Results and Discussion}

\bing{In this Section, we compute the values for $T_{ni}$ and $S_c$ for two models of Lebwohl-Lasher type, using the formulas we derived for the matched coefficients. For simplicity, we assume that the lattice is fully occupied (i.e., $\lambda \rightarrow \infty$) for both models. The first model (``model I") is a continuum version of the original model considered by Ref.~\cite{lebwohl-lasher1972}, in which the sum runs over nearest neighbours. The corresponding continuum representation of the nematogen interaction kernel $v$ is given by
\be
\label{eq:LLcont}
v(\rv - \rv') = \gamma a^3 \beta J \delta(\rv-\rv'), 
\ee
where $\gamma$ is the coordination number of each nematogen site. For a cubic lattice, $\gamma = 6$. A derivation of the above kernel is given in App.~\ref{app:LL}. The corresponding value of $\kappa$ is given by $\kappa = \gamma \beta J$.} 

\bing{The second model (``model II") that we consider is a modified version of the LL model, with allowance being made for interactions of finite (albeit short) range. In this model $v(\rv)$ has a Gaussian decaying form, i.e., 
\be
\label{eq:v-gaussian}
v(\rv) = \beta J \, {\rm e}^{-\frac{|\rv|^2}{2b^2}}, 
\ee
where $b$ sets the range of the nematogen interaction. For this potential, we have that $\kappa = (2\pi)^{3/2} \beta J (b/a)^3$.}

\bing{For model I, the matched coefficients are given by 
\begin{subequations}
\label{eq:matched-nn-full}
\ba
A &\!=\!& 6 \Big( 1 - \frac{12 \beta J}{15} \Big) J, 
\\
B &\!=\!& 864\beta^2 J^3/105,
\\
C &\!=\!& 33696 \beta^3 J^4/14175,
\\
L_1 &\!=\!& L_2 = 0
\ea
\end{subequations} 
From these values and Eq.~(\ref{eq:T*}), we find that $T^* = 12J/15k_{{\rm B}}$, $T_{ni} \approx 0.98 J/k_{{\rm B}}$, and $S_c \approx 0.85$.} 

\bing{For model II, we consider the case where $b = a$. The matched coefficients are then given by 
\begin{subequations}
\label{eq:occupy}
\ba
A &\!=\!& (2\pi)^{3/2} \left( 1 - \frac{2 (2\pi)^{3/2}\beta J}{15} \right) J,
\\
B &\!=\!& \frac{4(2\pi)^{9/2}}{105} \beta^2 J^3,
\\
C &\!=\!& \frac{1664 \pi^6}{14175} \beta^3 J^4,
\\
L_1 &\!=\!& 2 (2\pi)^{3/2} \left( \frac{(2\pi)^{3/2} \beta J}{15} 
- \frac{1}{4} \right) J \, a^2,
\\
L_2 &\!=\!& 0
\ea
\end{subequations}
From these values, we find that $T_{ni} \approx 2.6 J/k_{{\rm B}}$ and $S_c \approx 0.85$.} 

\bing{From the results obtained for the two models, we see that a lower transition temperature is predicted for the case of continuum nearest-neighbor interactions than for the case of finite-ranged interactions. This is physically reasonable, as an interaction of longer range implies that the nematogens are correlated over larger distances, whose correlations would require a correspondingly higher temperature to become disrupted. From the results of previous Monte Carlo (MC) simulations performed on the discrete, nearest-neighbor LL model~(see, e.g., Ref.~\cite{shekhar2012}), it is known that $T_{ni} \approx 1.11 J/k_{{\rm B}}$, which is distinct from the two values predicted by models I and II. The difference arises because the models are not identical with the (discrete) LL model. The value of $T_{ni}$ found by MC simulation falls between the values predicted by models I and II; this is so, because the interaction range in the discrete, nearest-neighbor LL model is larger than that in model I (which is a local approximation) but smaller than that in model II (in which the interaction can also be felt by next-nearest neighbors). In spite of the distinction, model I predicts that $T_{ni} \approx 0.98 J/k_{{\rm B}}$, which is not very far from the value found by MC simulation~\cite{shekhar2012}. Interestingly, both models I and II also predict approximately the same value for $S_c$. This value ($S_c \approx 0.85$) is larger than the value of 0.43 predicted by the Maier-Saupe theory and smaller than the value of 0.89 predicted by the Onsager theory~\cite{chandrasekhar,degennes-prost}, but close to the approximate value of 0.82 found by MC simulation~\cite{shekhar2012}.} 

\bing{In Eqs.~(\ref{eq:matched}), it may appear that the Landau-de Gennes terms do not vanish in the formal limit that $T \rightarrow 0$ where one would have expected them to, originating as they do from the entropy of nematogens~\cite{katriel1986,luckhurst2012}. To probe the $T\rightarrow 0$ limit, we would have to go back to Eq.~(\ref{eq:HeffQ}), which contains the exact log trace term for the entropic contribution; this term has indeed the correct prefactor of $T$. The LdG free energy is an expansion of the log trace term for temperatures near $T_{ni}$ (from the high-temperature side), and correspondingly the values of $T$ in the matched LdG coefficients have to be approximated to $T_{ni}$. On the other hand, if we take the formal limit $T \rightarrow \infty$ we find that the coefficients $B$ and $C$ vanish, leaving us only with terms of quadratic order, which is physically appropriate for the isotropic phase.} 

\bing{The original LL model~\cite{lebwohl-lasher1972} is one in which the translational degrees of freedom of the nematogens are frozen out. This is evident in the assumed full occupation of the lattice in which nematogens are not free to change positions. On the other hand, by allowing for partial occupation (i.e., $\phi < 1$) whilst constraining the total number of nematogens to be constant, the modified Lebwohl-Lasher model can in principle also account for the translational freedoms of nematogens that were unaccounted for in the original LL model.  } 

\section{Conclusion}

In this Paper, we have formulated a field-theoretic approach to determining the coefficients of the Landau-de Gennes theory \bing{for thermotropic uniaxial nematics}. As we have shown in our Paper, such an approach offers a systematic and relatively straightforward recipe for calculating all the phenomenological coefficients (including those of the gradient terms). \bing{We have applied this approach to the microscopic, Lebwohl-Lasher model, approximating the nematogen-nematogen interaction $v(\rv)$ by potentials with two distinct forms: a continuum version of the nearest-neighbor interaction (model I) and a Gaussian exponential-decaying form (model II). To carry out the sum over fluctuations, we have put the nematogens onto a cubic lattice, with each lattice cell being able to be occupied by at most one nematogen. Via this model we found that for the quartic order truncation of the LdG expansion to be stable, the fraction of occupied lattice cells has to be greater than $50/63$.} 

\bing{We have also derived formulas that express the coefficients $A$, $B$, $C$, $L_1$ and $L_2$ in terms of the parameters of the modified Lebwohl-Lasher model. By working with the two model potentials and assuming that the lattice is fully packed, we have obtained explicit values for the coefficients, and used them to deduce values for the isotropic-nematic transition temperature and the order parameter discontinuity. We found that for model I, $T_{ni} \approx 0.98 J/k_{{\rm B}}$ and $S_c \approx 0.85$, while for model II, $T_{ni} \approx 2.6 J/k_{{\rm B}}$ and $S_c \approx 0.85$. These predictions (coming from a continuum approximation) can be compared with the values for the nearest-neighbor Lebwohl-Lasher model found by MC simulations, viz., $T_{ni} \approx 1.11 J/k_{{\rm B}}$ and $S_c \approx 0.82$.} 

The systematic character of our field-theoretic approach implies that it can be straightforwardly extended to other liquid crystal systems, such as biaxial nematics, discoids, cholesterics and cholesteric blue phases. \bing{For biaxial nematics, one would have to expand the log trace term in the effective Hamiltonian to sixth order in $Q$~\cite{allender-longa2008}. This can be used to study the possibility of fluctuation-induced biaxial nematics, for negative values of $C$.} Our approach can in principle also be generalized to more complicated interaction potentials, for example, those involving electrostatic and Kirkwood-Shumaker interactions (for stiff rod-like peptides). 

\section{Acknowledgment}
\bing{The author thanks Dr Apala Majumdar for stimulating discussions and interest in the work, and also acknowledges the Accelerating International Collaboration Grant awarded by the University of Bath Internationalization Department to Dr Majumdar, which facilitated scientific exchanges between Nanyang Technological University and the University of Bath, including participation in a workshop devoted to the theme, ``Common Research Themes in Mathematics and Physics", organized by the Department of Mathematical Sciences and the Institute for Mathematical Innovation at the University of Bath. He also thanks the two anonymous referees for constructive comments.}

\appendix

\section{Expansion of the log trace term in the effective Hamiltonian}
\label{app:Hexpansion}

In this Appendix, we perform an expansion of the log trace term in $H_{{\rm eff}}$ in powers of $Q$, assuming that $Q$ is small. Using the shorthand notation $\sum_{\rv'} v_{\rv'\rv} \equiv \int d^3\rv' v(\rv'-\rv)/a^3$, we expand the logarithm to quartic order in $Q$: 
\ba
\label{eq:nonlocal}
&&\ln \! \big[ 
1 + \lambda \big\langle 
{\rm e}^{\sum_{\rv'} v_{\rv'\rv}\tau_{ab} Q_{ab}(\rv')}
\big\rangle_{n}
\big]
\\
&\approx& 
\ln \big[
1+ \lambda 
+ \lambda \langle \tau_{ab} \rangle_n \sum_{\rv'} v_{\rv'\rv} Q_{ab}(\rv')
\nonumber\\
&&\quad
+
\frac{\lambda}{2} 
\langle \tau_{ab}\tau_{cd} \rangle_n 
\sum_{\rv',\rv''} v_{\rv'\rv} v_{\rv''\rv}
Q_{ab}(\rv') Q_{cd}(\rv'')
\nonumber\\
&&\quad +
\frac{\lambda}{6}
\langle \tau_{ab}  \tau_{cd}  \tau_{ef} \rangle_n 
\!\! \sum_{\rv_1,\rv_2,\rv_3} \!\! 
v_{\rv_1\rv} v_{\rv_2\rv} v_{\rv_3\rv}
\nonumber\\
&&\qquad\times
Q_{ab}(\rv_1) Q_{cd}(\rv_2) Q_{ef}(\rv_3)
\nonumber\\
&&\quad +
\frac{\lambda}{24}
\langle \tau_{ab} \tau_{cd} \tau_{ef} \tau_{gh} \rangle_n 
\!\!\!\! \sum_{\rv_1,\rv_2,\rv_3,\rv_4} \!\!\!\! 
v_{\rv_1\rv} v_{\rv_2\rv} v_{\rv_3\rv} v_{\rv_4\rv}
\nonumber\\
&&\qquad\times
Q_{ab}(\rv_1) Q_{cd}(\rv_2) Q_{ef}(\rv_3) Q_{gh}(\rv_4) 
\big]
\nonumber\\
&\approx&
\frac{\lambda \langle \tau_{ab}\tau_{cd} \rangle_n}{2 (1+\lambda)} 
\sum_{\rv',\rv''} v_{\rv'\rv} v_{\rv''\rv}
Q_{ab}(\rv') Q_{cd}(\rv'')
\nonumber\\
&&\quad 
+
\frac{\lambda \langle \tau_{ab}  \tau_{cd}  \tau_{ef} \rangle_n}{6 (1+\lambda)}
\!\! \sum_{\rv_1,\rv_2,\rv_3} \!\! 
v_{\rv_1\rv} v_{\rv_2\rv} v_{\rv_3\rv}
\nonumber\\
&&\qquad\times
Q_{ab}(\rv_1) Q_{cd}(\rv_2) Q_{ef}(\rv_3)
\nonumber\\
&& +
\left[ 
\frac{\lambda \langle \tau_{ab} \tau_{cd} \tau_{ef} \tau_{gh} \rangle_n}{24 (1+\lambda)}
-
\frac{\lambda^2 \langle \tau_{ab} \tau_{cd} \rangle_n \langle \tau_{ef} \tau_{gh} \rangle_n}{8 (1+\lambda)^2}
\right]
 \nonumber\\
 &&\times \!\!\!\!
 \!\!\!\! \sum_{\rv_1,\rv_2,\rv_3,\rv_4} \!\!\!\!\!\! 
v_{\rv_1\rv} v_{\rv_2\rv} v_{\rv_3\rv} v_{\rv_4\rv} \, 
Q_{ab}(\rv_1) Q_{cd}(\rv_2) Q_{ef}(\rv_3) Q_{gh}(\rv_4) 
\nonumber\\
 &&+ \, {\rm const}
\nonumber
\ea
In the above, a term linear in $Q$ vanishes as $\langle \tau_{ab} \rangle_n = 0$. 
To make further progress, we make a local approximation to the cubic and quartic terms, which consists of expressing $\rv_\alpha$ (where $\alpha=1,2,3$ in the cubic term and $\alpha=1,2,3,4$ in the quartic term) in Eq.~(\ref{eq:nonlocal}) in terms of a centre coordinate $\rv$ and relative coordinates $\Rv_\alpha$, viz., $\rv_\alpha = \rv+\Rv_\alpha$, and taking $\rv_\alpha \approx \rv$. Taking the local approximation is reasonable considering that near the transition, $Q$ would be small, and nonlocal variations of $Q$ introduce higher order corrections that can  be disregarded in terms of order higher than quadratic.

\section{Equivalence of the expectation value of the auxiliary field $\langle Q \rangle$ and the ensemble-averaged collective field $\langle q \rangle$}
\label{app:Q}

Here we show that the expectation value of $Q$, obtained by averaging over $Q$ with a Boltzmann weight $e^{-\beta H_{{\rm eff}}}$, coincides with the ensemble average of $q$.
To see their equivalence, let us append an external source term to $Z$ in Eq.~(\ref{eq:Z}), viz., 
\ba
\label{eq:anne}
Z[h] &\!=\!& \prod_{j=1}^{N} \sum_{s_j=0,1} \! {\rm e}^{\beta \mu \sum_j s_j} \! 
\\
&&\quad \times \, 
\Big\langle 
{\rm e}^{\frac{1}{2} \!\int\! \frac{d^3\rv}{a^3} \! \int \! \frac{d^3\rv'}{a^3} \, 
q_{ab}(\rv) v(\rv - \rv') q_{ab}(\rv') + \int \! \frac{d^3\rv}{a^3} \, h_{ab} q_{ab}}
\Big\rangle_{n_j}
\nonumber
\ea
whence the ensemble average of $q$ is given by
\be
\label{eq:Qe1}
\langle q_{ab}(\rv) \rangle = \left.\frac{1}{Z}\frac{\partial Z}{\partial h_{ab}(\rv)}\right|_{h\rightarrow 0}
\ee
To make a connection to the expectation value of $Q$, we perform a HS transformation on $Z[h]$ in Eq.~(\ref{eq:anne}): 
\ba
&&Z[h]
\\
&\!\!=\!\!&
\prod_{j=1}^{N} \sum_{s_j=0,1} \! {\rm e}^{\beta \mu \sum_j s_j - \frac{1}{2}  \!\int\! \frac{d^3\rv}{a^3} \! \int \! \frac{d^3\rv'}{a^3} \, h_{ab}(\rv) v^{-1}(\rv - \rv') h_{ab}(\rv')}
\nonumber\\ 
&&\quad \times \, 
\Big\langle 
{\rm e}^{\frac{1}{2} \!\int\! \frac{d^3\rv}{a^3} \! \int \! \frac{d^3\rv'}{a^3} \, 
(q_{ab} + \int \! v^{-1} h_{ab}) v (q_{ab} + \int \! v^{-1} h_{ab})}
\Big\rangle_{n_j}
\nonumber\\ 
&\!\!=\!\!& \int\!\mathcal{D}Q \, 
{\rm e}^{-\frac{1}{2} \!\int\! \frac{d^3\rv}{a^3} \! \int \! \frac{d^3\rv'}{a^3} \, 
Q_{ab}(\rv) v(\rv - \rv') Q_{ab}(\rv')}
\nonumber\\
&&\times
\prod_{j=1}^{N} 
\sum_{s_j=0,1} \! 
{\rm e}^{\beta \mu \sum_{j} s_j} 
\Big\langle {\rm e}^{\int \!\frac{d^3\rv}{a^3} \! \int \!\frac{d^3\rv'}{a^3} q_{ab}(\rv) v(\rv-\rv') Q_{ab}(\rv')}
\nonumber\\
&&\qquad\qquad\times\,
{\rm e}^{\int\!\frac{d^3\rv}{a^3} h_{ab} Q_{ab}}
\Big\rangle_{n_j}
\nonumber
\ea
From this equation, we obtain the relation
\be
\label{eq:Qe2}
 \left.\frac{1}{Z}\frac{\partial Z}{\partial h_{ab}(\rv)}\right|_{h\rightarrow 0}
= \langle Q_{ab}(\rv) \rangle.
\ee
Comparing Eqs.~(\ref{eq:Qe1}) and (\ref{eq:Qe2}), we have the equality 
\be
\label{eq:dualrelation}
\langle q_{ab}(\rv) \rangle = \langle Q_{ab}(\rv) \rangle. 
\ee

\section{Constrained averages over $S^2$}
\label{app:tensors}

As we see from Eq.~(\ref{eq:HeffQ}), the effective Hamiltonian contains terms of the form $\langle
e^{\tau_{\alpha\beta}Q_{\alpha\beta}(\rv)}
\rangle_\nv$, where $\langle \ldots \rangle_\nv$ is an average over the unit 2-sphere $S^2$ (i.e., the set of points mapped out by all possible orientations of a unit vector, $\nv$). Close to the isotropic-nematic transition point, we expect $Q$ to be small, and thus an expansion in powers of $Q$ can be performed. To match our effective Hamiltonian to the Landau-de Gennes free energy, we have to carry out the expansion to quartic order in $Q$. Such terms involve orientation averages over products of $\tau$, and the highest order of such products in the quartic-order expansion is obviously of quartic order. The evaluation of such orientation averages is equivalent to the evaluation of orientation averages over products of ${\bf n}$. To quartic order in $Q$, we find that the orientation averages yield isotropic tensors of order 2, 4, 6 and 8. The results are given below: 
\ba
\langle 1 \rangle_{\bf n} &\!\!=\!\!& 1 
\\
\langle n_{\alpha} n_{\beta} \rangle_{\bf n} &\!\!=\!\!& \frac{1}{3}\delta_{\alpha\beta} 
\\
\langle n_{\alpha} n_{\beta} n_{\gamma} n_{\delta} \rangle_{\bf n}
&\!\!=\!\!& 
\frac{1}{15}(\delta_{\alpha\beta}\delta_{\gamma\delta} + \delta_{\alpha\gamma}\delta_{\beta\delta} + \delta_{\alpha\delta}\delta_{\beta\gamma}) 
\\
\langle n_{a} n_{b} n_{c} n_{d} n_e n_f \rangle_{\bf n}
&\!\!=\!\!&
\frac{1}{105} \big( 
\delta_{ab} (\delta_{cd}\delta_{ef} + \delta_{ce}\delta_{df} + \delta_{cf}\delta_{de})
\nonumber\\
&&\quad\,\,\,+
\delta_{ac} (\delta_{bd}\delta_{ef} + \delta_{be}\delta_{df} + \delta_{bf}\delta_{de})
\nonumber\\
&&\quad\,\,\,+
\delta_{ad} (\delta_{bc}\delta_{ef} + \delta_{be}\delta_{cf} + \delta_{bf}\delta_{ce})
\nonumber\\
&&\quad\,\,\,+
\delta_{ae} (\delta_{bc}\delta_{df} + \delta_{bd}\delta_{cf} + \delta_{bf}\delta_{cd})
\nonumber\\
&&\quad\,\,\,+
\delta_{af} (\delta_{bc}\delta_{de} + \delta_{bd}\delta_{ce} + \delta_{be}\delta_{cd})
\big)   
\nonumber\\
\ea
The second equation above implies that $\langle \tau_{\alpha\beta} \rangle_{\bf n} = 0$.
In the free energy expansion, when we contract $\tau_{ab}$ with tensors $Q_{ab}$ in the free energy expansion, only traceless contributions will be picked up, and thus we can ignore terms involving $\langle n_{a} n_{b} \rangle_{\bf n}$, and we effectively have that 
\ba
&&\langle n_{a} n_{b} n_{c} n_{d} \rangle_{\bf n}
\rightarrow 
\frac{1}{15}(\delta_{ac}\delta_{bd} + \delta_{ad}\delta_{bc}), 
\\
&&\langle n_{a} n_{b} n_{c} n_{d} n_e n_f \rangle_{\bf n}
\\
&\rightarrow& 
\frac{1}{105} \big( 
\delta_{ac} (\delta_{be}\delta_{df} + \delta_{bf}\delta_{de})
+
\delta_{ad} (\delta_{be}\delta_{cf} + \delta_{bf}\delta_{ce})
\nonumber\\
&&\quad\,\,\,+
\delta_{ae} (\delta_{bc}\delta_{df} + \delta_{bd}\delta_{cf})
+
\delta_{af} (\delta_{bc}\delta_{de} + \delta_{bd}\delta_{ce})
\big). 
\nonumber
\ea
Using the above results, we calculate the following quantity which is relevant for the coefficient of the quadratic term in the free energy expansion of $Q$: 
\ba
&&\langle \tau_{ab}\tau_{cd} \rangle_{\bf n} 
\\
&=& 
\langle (n_a n_b - \frac{1}{3}\delta_{ab})(n_c n_d - \frac{1}{3}\delta_{cd}) \rangle_{\bf n}
\nonumber\\
&=& 
\langle 
n_a n_b n_c n_d - \frac{1}{3}\delta_{ab} n_c n_d - \frac{1}{3}\delta_{cd} n_a n_b + \frac{1}{9} \delta_{ab}\delta_{cd} 
\rangle_{\bf n}
\nonumber\\
&=& 
-\frac{2}{45} \delta_{ab}\delta_{cd} + \frac{1}{15} (\delta_{ac}\delta_{bd} + \delta_{ad}\delta_{bc}), 
\nonumber
\ea
For the cubic order term in the free energy expansion, we need the value of $\langle \tau_{ab} \tau_{cd} \tau_{ef} \rangle_{\nv}$:
\ba
&&\langle  \tau_{ab} \tau_{cd} \tau_{ef} \rangle_{\bf n}
\\
&\!\!=\!\!&
\frac{1}{105} \big( 
\delta_{ac} ( \delta_{be}\delta_{df} + \delta_{bf}\delta_{de})
+
\delta_{ad} ( \delta_{be}\delta_{cf} + \delta_{bf}\delta_{ce})
\nonumber\\
&&\quad\,\,\,+
\delta_{ae} (\delta_{bc}\delta_{df} + \delta_{bd}\delta_{cf} )
+
\delta_{af} (\delta_{bc}\delta_{de} + \delta_{bd}\delta_{ce} )
\big)   
\nonumber
\ea
The quartic term in the free energy expansion is a connected cumulant term consisting of two contributions. The first contribution is straightforward to calculate:
\ba
&&\langle \tau_{ab}\tau_{cd} \rangle_{\bf n} \langle \tau_{ef}\tau_{gh} \rangle_{\bf n}
\\
&=& 
\frac{4}{2025}\delta_{ab}\delta_{cd}\delta_{ef}\delta_{gh}
- \frac{2}{225} 
(\delta_{ab}\delta_{cd}\delta_{eg}\delta_{fh} 
+ \delta_{ab}\delta_{cd}\delta_{eh}\delta_{fg}
\nonumber\\
&&\qquad+ 
\delta_{ac}\delta_{bd}\delta_{ef}\delta_{gh} 
+ \delta_{ad}\delta_{bc}\delta_{ef}\delta_{gh})
\nonumber\\
&&
+ \frac{1}{225} (\delta_{ac}\delta_{bd}\delta_{eg}\delta_{fh} 
+ \delta_{ac}\delta_{bd}\delta_{eh}\delta_{fg}
+ \delta_{ad}\delta_{bc}\delta_{eg}\delta_{fh}
\nonumber\\
&&\qquad 
+ \delta_{ad}\delta_{bc}\delta_{eh}\delta_{fg})
\nonumber\\
&=&
\frac{1}{225} (\delta_{ac}\delta_{bd}\delta_{eg}\delta_{fh} 
+ \delta_{ac}\delta_{bd}\delta_{eh}\delta_{fg}
+ \delta_{ad}\delta_{bc}\delta_{eg}\delta_{fh} 
\nonumber\\
&&\qquad 
+ \delta_{ad}\delta_{bc}\delta_{eh}\delta_{fg}).
\nonumber
\ea

To obtain the second contribution, we have to consider the following rank-eight isotropic tensor:
\begin{widetext}
\ba
&&\langle n_{a} n_{b} n_{c} n_{d} n_e n_f n_g n_h \rangle_{\bf n}
\\
&=&
\frac{1}{945}
\big\{
\delta_{ab}
\big[
\delta_{cd} (\delta_{ef}\delta_{gh} + \delta_{eg}\delta_{fh} + \delta_{eh}\delta_{fg})
+
\delta_{ce} (\delta_{df}\delta_{gh} + \delta_{dg}\delta_{fh} + \delta_{dh}\delta_{fg})
+
\delta_{cf} (\delta_{de}\delta_{gh} + \delta_{dg}\delta_{eh} + \delta_{dh}\delta_{eg})
\nonumber\\
&&\quad\qquad\,
+
\delta_{cg} (\delta_{de}\delta_{fh} + \delta_{df}\delta_{eh} + \delta_{dh}\delta_{ef})
+
\delta_{ch} (\delta_{de}\delta_{fg} + \delta_{df}\delta_{eg} + \delta_{dg}\delta_{ef})
\big]
\nonumber\\
&&\quad\,\,\,
+
\delta_{ac}
\big[ 
\delta_{bd} (\delta_{ef}\delta_{gh} + \delta_{eg}\delta_{fh} + \delta_{eh}\delta_{fg})
+
\delta_{be} (\delta_{df}\delta_{gh} + \delta_{dg}\delta_{fh} + \delta_{dh}\delta_{fg})
+
\delta_{bf} (\delta_{de}\delta_{gh} + \delta_{dg}\delta_{eh} + \delta_{dh}\delta_{eg})
\nonumber\\
&&\quad\qquad\,
+
\delta_{bg} (\delta_{de}\delta_{fh} + \delta_{df}\delta_{eh} + \delta_{dh}\delta_{ef})
+
\delta_{bh} (\delta_{de}\delta_{fg} + \delta_{df}\delta_{eg} + \delta_{dg}\delta_{ef})
\big]
\nonumber\\
&&\quad\,\,\,
+
\delta_{ad}
\big[ 
\delta_{bc} (\delta_{ef}\delta_{gh} + \delta_{eg}\delta_{fh} + \delta_{eh}\delta_{fg})
+
\delta_{be} (\delta_{cf}\delta_{gh} + \delta_{cg}\delta_{fh} + \delta_{ch}\delta_{fg})
+
\delta_{bf} (\delta_{ce}\delta_{gh} + \delta_{cg}\delta_{eh} + \delta_{ch}\delta_{eg})
\nonumber\\
&&\quad\qquad\,
+
\delta_{bg} (\delta_{ce}\delta_{fh} + \delta_{cf}\delta_{eh} + \delta_{ch}\delta_{ef})
+
\delta_{bh} (\delta_{ce}\delta_{fg} + \delta_{cf}\delta_{eg} + \delta_{cg}\delta_{ef})
\big]
\nonumber\\
&&\quad\,\,\,+
\delta_{ae}
\big[ 
\delta_{bc} (\delta_{df}\delta_{gh} + \delta_{dg}\delta_{fh} + \delta_{dh}\delta_{fg})
+
\delta_{bd} (\delta_{cf}\delta_{gh} + \delta_{cg}\delta_{fh} + \delta_{ch}\delta_{fg})
+
\delta_{bf} (\delta_{cd}\delta_{gh} + \delta_{cg}\delta_{dh} + \delta_{ch}\delta_{dg})
\nonumber\\
&&\quad\qquad\,
+
\delta_{bg} (\delta_{cd}\delta_{fh} + \delta_{cf}\delta_{dh} + \delta_{ch}\delta_{df})
+
\delta_{bh} (\delta_{cd}\delta_{fg} + \delta_{cf}\delta_{dg} + \delta_{cg}\delta_{df})
\big]
\nonumber\\
&&\quad\,\,\,+
\delta_{af}
\big[ 
\delta_{bc} (\delta_{de}\delta_{gh} + \delta_{dg}\delta_{eh} + \delta_{dh}\delta_{eg})
+
\delta_{bd} (\delta_{ce}\delta_{gh} + \delta_{cg}\delta_{eh} + \delta_{ch}\delta_{eg})
+
\delta_{be} (\delta_{cd}\delta_{gh} + \delta_{cg}\delta_{dh} + \delta_{ch}\delta_{dg})
\nonumber\\
&&\quad\qquad\,
+
\delta_{bg} (\delta_{cd}\delta_{eh} + \delta_{ce}\delta_{dh} + \delta_{ch}\delta_{de})
+
\delta_{bh} (\delta_{cd}\delta_{eg} + \delta_{ce}\delta_{dg} + \delta_{cg}\delta_{de})
\big]
\nonumber\\
&&\quad\,\,\,+
\delta_{ag}
\big[ 
\delta_{bc} (\delta_{de}\delta_{fh} + \delta_{df}\delta_{eh} + \delta_{dh}\delta_{ef})
+
\delta_{bd} (\delta_{ce}\delta_{fh} + \delta_{cf}\delta_{eh} + \delta_{ch}\delta_{ef})
+
\delta_{be} (\delta_{cd}\delta_{fh} + \delta_{cf}\delta_{dh} + \delta_{ch}\delta_{df})
\nonumber\\
&&\quad\qquad\,
+
\delta_{bf} (\delta_{cd}\delta_{eh} + \delta_{ce}\delta_{dh} + \delta_{ch}\delta_{de})
+
\delta_{bh} (\delta_{cd}\delta_{ef} + \delta_{ce}\delta_{df} + \delta_{cf}\delta_{de})
\big]
\nonumber\\
&&\quad\,\,\,+
\delta_{ah}
\big[ 
\delta_{bc} (\delta_{de}\delta_{fg} + \delta_{df}\delta_{eg} + \delta_{dg}\delta_{ef})
+
\delta_{bd} (\delta_{ce}\delta_{fg} + \delta_{cf}\delta_{eg} + \delta_{cg}\delta_{ef})
+
\delta_{be} (\delta_{cd}\delta_{fg} + \delta_{cf}\delta_{dg} + \delta_{cg}\delta_{df})
\nonumber\\
&&\quad\qquad\,
+
\delta_{bf} (\delta_{cd}\delta_{eg} + \delta_{ce}\delta_{dg} + \delta_{cg}\delta_{de})
+
\delta_{bg} (\delta_{cd}\delta_{ef} + \delta_{ce}\delta_{df} + \delta_{cf}\delta_{de})
\big]
\big\}
\ea
\end{widetext}
The above leads to the following useful result: 
\be
\langle \tau_{ab}\tau_{cd}\tau_{ef}\tau_{gh} \rangle_{\bf n} 
\rightarrow
\langle
n_a n_b n_c n_d n_e n_f n_g n_h
\rangle_{\bf n},   
\ee
so $\langle \tau_{ab}\tau_{cd}\tau_{ef}\tau_{gh} \rangle_{\bf n}$ effectively has only one contribution, the other contributions vanishing on account of the tracelessness of the $Q_{ab}$ tensors that are coupled to the $\tau_{ab}$ tensors. 

\section{Calculation of $I_1$ and $I_2$}
\label{app:calc}

Here we calculate the functions $I_1$ and $I_2$ defined in Eqs.~(\ref{eq:I1}) and (\ref{eq:I2}). 
The calculation can be facilitated by expressing $\rv = \cv + \Rv/2$, $\rv' = \cv - \Rv/2$ for the first quadratic term and $\rv' = \rv + \Rv_1$, $\rv'' = \rv + \Rv_2$ for the second, and we expand to quadratic order in the relative displacements $\Rv$, $\Rv_1$ and $\Rv_2$. $I_1$ becomes  
\ba
&&I_1 \equiv \frac{1}{2} 
\!\int \! \frac{d^3\rv}{a^3} \! \int \! \frac{d^3\rv'}{a^3} \, 
 v(\rv-\rv') Q_{ab}(\rv) Q_{ab}(\rv') 
\nonumber\\
&=&
\frac{1}{2}
\!\int \! \frac{d^3\cv}{a^3} \! \int \! \frac{d^3\Rv}{a^3} \, 
v(\Rv) Q_{ab}(\cv+\tfrac{1}{2}\Rv) Q_{ab}(\cv+\tfrac{1}{2}\Rv) 
\nonumber\\
&\approx&
\frac{1}{2}
\!\int \! \frac{d^3\cv}{a^3} \! \int \! \frac{d^3\Rv}{a^3} \, 
v(\Rv) 
\nonumber\\
&&\times 
(Q_{ab}(\cv) + \tfrac{1}{2}R_c \partial_c Q_{ab}(\cv) 
+ \tfrac{1}{8}R_c R_d \partial_c \partial_d Q_{ab}(\cv))
\nonumber\\
&&\times
(Q_{ab}(\cv) - \tfrac{1}{2}R_c \partial_c Q_{ab}(\cv) 
+ \tfrac{1}{8}R_e R_f \partial_e \partial_f Q_{ab}(\cv))
\nonumber\\
&\approx&
\frac{1}{2}
\!\int \! \frac{d^3\cv}{a^3} \! \int \! \frac{d^3\Rv}{a^3} \, 
v(\Rv) Q_{ab}(\cv) Q_{ab}(\cv)
\nonumber\\
&&+ 
\frac{1}{8}
\!\int \! \frac{d^3\cv}{a^3} \! \int \! \frac{d^3\Rv}{a^3} \, 
v(\Rv) \Big[ 
(R_c R_d \partial_c \partial_d Q_{ab}(\cv)) Q_{ab}(\cv)
\nonumber\\
&&\quad 
- 
R_c R_d \partial_c Q_{ab}(\cv) \partial_d Q_{ab}(\cv)
\Big]
\nonumber
\ea
On integrating the last term by parts, we obtain
\ba
&&I_1 = \frac{1}{2}
\!\int \! \frac{d^3\cv}{a^3} \! \int \! \frac{d^3\Rv}{a^3} \, 
v(\Rv) Q_{ab}(\cv) Q_{ab}(\cv)
\nonumber\\
&&- 
\frac{1}{4}
\!\int \! \frac{d^3\cv}{a^3} \! \int \! \frac{d^3\Rv}{a^3} \, 
(R_c R_d \, v(\Rv)) 
 \partial_c Q_{ab}(\cv) \partial_d Q_{ab}(\cv)
\nonumber\\
&& + \,\, {\text{surface term}}
\ea
For $I_2$ we have
\ba
&&I_2 \equiv 
\! \int \! \frac{d^3\rv}{a^3} 
\! \int \! \frac{d^3\rv'}{a^3}
\! \int \! \frac{d^3\rv''}{a^3}
v(\rv'-\rv) v(\rv''-\rv)
\nonumber\\
&&\quad \times \, 
Q_{ab}(\rv') Q_{ab}(\rv'') 
\nonumber\\
&=& 
\! \int \! \frac{d^3\rv}{a^3} 
\! \int \! \frac{d^3\Rv_1}{a^3}
\! \int \! \frac{d^3\Rv_2}{a^3}
v(\Rv_1) v(\Rv_2)
\nonumber\\
&&\quad \times \, 
Q_{ab}(\rv+\Rv_1) Q_{ab}(\rv+\Rv_2) 
\nonumber\\
&\approx& 
\! \int \! \frac{d^3\rv}{a^3} 
\! \int \! \frac{d^3\Rv_1}{a^3}
\! \int \! \frac{d^3\Rv_2}{a^3}
v(\Rv_1) v(\Rv_2)
\nonumber\\
&&\quad \times \, 
(Q_{ab}(\rv) + R_{1c}\partial_c Q_{ab}(\rv) + \tfrac{1}{2}R_{1c}R_{1d}\partial_c \partial_d Q_{ab}(\rv))
\nonumber\\
&&\quad\times 
(Q_{ab}(\rv) + R_{2e}\partial_e Q_{ab}(\rv) + \tfrac{1}{2}R_{2e}R_{2f}\partial_e \partial_f Q_{ab}(\rv))
\nonumber
\ea
We note that terms linear in $\Rv$ integrated over the isotropic potential $v(\Rv)$ vanish, which leads to
\ba
I_2
&\approx& 
\left( \int \! \frac{d^3\Rv}{a^3}
v(\Rv) \right)^2
\!\! \int \! \frac{d^3\rv}{a^3} 
Q_{ab}(\rv) Q_{ab}(\rv) 
\nonumber\\
&&
+
\! \int \! \frac{d^3\Rv_1}{a^3} 
v(\Rv_1) 
\! \int \! \frac{d^3\Rv_2}{a^3}
R_{2c} R_{2d} \, v(\Rv_2) 
\nonumber\\
&&\quad\times
\! \int \! \frac{d^3\rv}{a^3} 
(\partial_c \partial_d Q_{ab}(\rv))
Q_{ab}(\rv)
\nonumber\\
&\approx& 
\left( \int \! \frac{d^3\Rv}{a^3}
v(\Rv) \right)^2
\!\! \int \! \frac{d^3\rv}{a^3} 
Q_{ab}(\rv) Q_{ab}(\rv) 
\nonumber\\
&&
-
\! \int \! \frac{d^3\Rv_1}{a^3} 
v(\Rv_1) 
\! \int \! \frac{d^3\Rv_2}{a^3}
R_{2c} R_{2d} \, v(\Rv_2) 
\nonumber\\
&&\quad\times
\! \int \! \frac{d^3\rv}{a^3} 
\partial_c Q_{ab}(\rv)
\partial_d Q_{ab}(\rv)
\nonumber\\
&&\quad + \, {\text{surface term}}
\nonumber
\ea
On going to the last equality, we have performed a partial integration.

\section{Continuum version of nearest-neighbor LL model}
\label{app:LL}

\bing{
Here we derive the continuum version of the nearest-neighbor LL model (i.e., Eq.~(\ref{eq:LLcont})). We consider a cubic lattice where the $i$-th nematogen ($i=1,\ldots, N_{nem}$) occupies a site that has a position vector $\rv_i = x_i {\bf{e}}_x + y_i {\bf{e}}_y + z_i {\bf{e}}_z$. Here $\bf{e}_x$, $\bf{e}_y$ and $\bf{e}_z$ are unit vectors directed along the three mutually orthogonal axes of the cubic lattice. Let us consider a nematogen $j$, and the quantity $\widehat{\sum}_i v(\rv_i,\rv_j)Q(\rv_i)$, where $\widehat{\sum}_i$ denotes the sum over the nearest neighbors of $j$: 
\ba
&&\widehat{\sum}_i v(\rv_i,\rv_j)Q(\rv_i) 
\nonumber\\
&\!\equiv\!& \beta J \sum_i \, [ \delta_{x_j x_i} \delta_{y_j y_i} (\delta_{z_j, z_i+a} + \delta_{z_j, z_i-a})
\nonumber\\
&&+ \delta_{y_j y_i} \delta_{z_j z_i} (\delta_{x_j, x_i+a} + \delta_{x_j, x_i-a})
\nonumber\\
&&+ \delta_{z_j z_i} \delta_{x_j x_i} (\delta_{y_j, y_i+a} + \delta_{y_j, y_i-a}) ] Q(\rv_i), 
\ea
The sum on the right-hand side now runs over all values of $i$. In the continuum limit, the sites neighboring $j$ ``collapse" onto the site $j$, which leads to
\be
\widehat{\sum}_i v(\rv_i,\rv_j)Q(\rv_i) \rightarrow \int \! \frac{d^3\rv'}{a^3} (a^3\beta J) \gamma \delta(\rv-\rv') Q(\rv'),
\ee
where $\gamma = 6$ is the coordination number for a cubic lattice. We can thus extract the continuum limit of $v$, viz.,
\be
v(\rv-\rv') = \gamma \beta J a^3 \delta(\rv - \rv').
\ee
}


\begin{thebibliography}{99}
\bibitem{footnote1}
We can see this by using the representation $Q_{ab} = \left( 
\begin{array}{ccc}
0 & Q_1 & Q_2 \\
Q_1 & 0 & Q_3 \\
Q_2 & Q_3 & 0 
\end{array} 
\right)$, from which we obtain $\Tr Q^4 = 2(Q_1^2+Q_2^2+Q_3^2)^2$ and $(\Tr Q^2)^2 = 4(Q_1^2+Q_2^2+Q_3^2)^2$.
\end{thebibliography}
\end{document}